\documentclass{iau}
\usepackage{graphicx}
\usepackage{amsmath}

\usepackage{url}

\usepackage{xspace}

\def\araa{ARA\&A}
\def\apj{ApJ}
\def\apjl{ApJ}
\def\apjs{ApJS}

\def\aap{A\&A}

\def\aaps{A\&AS}

\def\mnras{MNRAS}

\def\pasj{PASJ}

\def\nat{Nature}

\def\rmxaa{RMxAA}

\newcommand{\maihem}{\textsc{maihem}\xspace}
\newcommand{\flash}{\textsc{flash}\xspace}
\newcommand{\cloudy}{\textsc{cloudy}\xspace}

\newcommand{\ionic}[2]{#1$\,${\scshape{#2}}\xspace}

\usepackage{natbib}
\usepackage{color,hyperref}
\hypersetup{breaklinks=true,colorlinks=true,
            linkcolor=blue,urlcolor=blue,anchorcolor=blue,citecolor=blue,filecolor=blue,
            bookmarks=true, bookmarksopen=true, bookmarksopenlevel=1}  
            
\title[Superwinds with Time-evolving Stellar Feedback] 
{Numerical Modeling of Galactic Superwinds with Time-evolving Stellar Feedback}

\author[A. Danehkar, M. S. Oey, \& W. J. Gray]   
{A. Danehkar$^1$, M. S. Oey$^2$, and W. J. Gray}

\affiliation{$^1$\,Eureka Scientific, 2452 Delmer Street, Suite 100, Oakland, CA 94602-3017, USA
\\ [\affilskip]$^2$\,Department of Astronomy, University of Michigan, Ann Arbor, MI 48109, USA
\\email: {\tt \href{mailto:danehkar@eurekasci.com}{danehkar@eurekasci.com}}}

\pubyear{2023}
\volume{370}  
\pagerange{217-222}
\doi{\href{https://doi.org/10.1017/S1743921323000066}{10.1017/S1743921323000066}}
\jname{Winds of Stars and Exoplanets}
\editors{A.~A. Vidotto, L.~Fossati \& J.~Vink, eds.}
\begin{document}

\maketitle

\begin{abstract}
Mass-loss and radiation feedback from evolving massive stars produce galactic-scale superwinds, sometimes surrounded by pressure-driven bubbles. Using the time-dependent stellar population typically seen in star-forming regions, we conduct hydrodynamic simulations of a starburst-driven superwind model coupled with radiative efficiency rates to investigate the formation of radiative cooling superwinds and bubbles. Our numerical simulations depict the parameter space where radiative cooling superwinds with or without bubbles occur. Moreover, we employ the physical properties and time-dependent ionization states to predict emission line profiles under the assumption of collisional ionization and non-equilibrium ionization caused by wind thermal feedback in addition to photoionization created by the radiation background. We see the dependence of non-equilibrium ionization structures on the time-evolving ionizing source, leading to a deviation from collisional ionization in radiative cooling wind regions over time.
\end{abstract}

\firstsection 

\section{Introduction}

Galactic-scale superwinds emerging from star-forming galaxies have commonly been seen in 
several multiwavelength observations \citep[see e.g.,][]{Heckman1990,Rupke2005,Veilleux2005}, 
which are sometimes accompanied by large-scale bubbles \citep[e.g.,][]{Veilleux1994,Sakamoto2006,Tsai2009}.
Moreover, some observations of compact starburst regions pointed to unexpected cooling and suppressed superwinds \citep{Oey2017,Turner2017,Jaskot2017}, which could not be completely explained by the standard model based on the adiabatic assumption \citep{Chevalier1985,Canto2000}. However, these phenomena could be related to heat being lost through radiation. In particular, semi-analytical studies of superwind models with radiative cooling found that the wind temperature could deviate from the adiabatic result depending on the stellar mass-loss rate and wind velocity \citep{Silich2004,Tenorio-Tagle2005}, which has been confirmed by recent hydrodynamic simulations \citep{Gray2019a,Danehkar2021}.

While mechanical feedback from massive OB stars could create pressure-driven bubbles,
some bubbles seem to expand more slowly than predicted \citep[see e.g.,][]{Brown1995,Oey1996}.
This could be explained by the time-dependent stellar feedback from evolving massive stars, which undergo stellar evolution, i.e., OB star $\rightarrow$ RSG/LBV $\rightarrow$ WR star.
Recently, \citet{Danehkar2021,Danehkar2022} implemented hydrodynamic simulations and photoionization models of superwinds for different wind parameters using the stellar feedback from a stellar population at a fixed age of 1\,Myr. However,
taking the entire mass-loss history of massive stars can significantly change the theoretical predictions of pressure-driven bubbles over a timescale larger than 1\,Myr \citep[see e.g.,][]{Oey2004,Krause2013}. Moreover, a time-evolving photoionizing source could modify non-equilibrium photoionization predictions, which primarily rely on time-dependent ionization states made by hydrodynamic simulations using the radiation background.

\section{Numerical Modeling of Galactic Superwinds}

We conducted hydrodynamic simulations of a spherically symmetric superwind model coupled with the radiative efficiency rates using the non-equilibrium chemistry package \maihem \citep{Gray2019} built on the adaptive mesh hydrodynamics code \flash \citep{Fryxell2000}, which solves the following fluid equations:
\begin{align}
\frac{d \rho}{d t}+\frac{1}{r^2} \frac{d}{dr} \left( \rho u r^2 \right)  = & q_{m}(t),  \label{eq_1} \\ 
\frac{d \rho u}{d t}+\rho u \frac{d u}{dr} + \frac{d P}{dr}  = &- q_{m}(t) u, \label{eq_2} \\ 
\frac{d {\rho E}}{d t} + \frac{1}{r^2} \frac{d}{dr} \left[ \rho u r^2  \left( \frac{u^{2}}{2} +\frac{\gamma}{\gamma -1} \frac{P}{\rho}  \right) \right] = & \displaystyle\sum_{i}^{}  n_i  \Gamma_i(t)-\displaystyle\sum_{i}^{}  n_i n_e \Lambda_i+ q_{e}(t), \label{eq_3} \\
\frac{1}{n_{\rm e}}\frac{d n_{i}}{d t}  =& n_{i+1}\alpha_{i+1} - n_{i}\alpha_{i} + n_{i-1} S_{i-1} \notag \\ 
& - n_{i}S_{i} + \frac{1}{n_{\rm e}} n_{i-1} \zeta_{i-1} (t) - \frac{1}{n_{\rm e}} n_{i} \zeta_{i} (t),  \label{eq_4}
\end{align}
where $r$, $\rho$, $u$, $P$, and $E$  are the radius,  density, velocity, pressure, and energy per mass of the fluid, respectively, $\gamma=5/3$ is the specific heat ratio, $q_{m}(t) = \dot{M}(t) / (\frac{4}{3} \pi R^3_{\rm sc})$ and $q_{e} (t)=  \frac{1}{2} \dot{M}(t) V_{\infty}(t)^2  / (\frac{4}{3} \pi R^3_{\rm sc})$ are the time-dependent mass and energy injection rates per volume according to the mass-loss rate $\dot{M}(t)$ and wind velocity $V_{\infty}(t)$ of the time-evolving stellar population produced by Starburst99, respectively, $n_i$ is the ion densities,  $n_e$ the electron density, $\Lambda_i$ the radiative cooling rates for the given temperature $T$ derived from the cooling atomic data \citep{Gnat2012}, $\Gamma_i (t) = \int^{\infty}_{\nu_{0,i}} (4 \pi J_{\nu} (t)/\nu) (\nu-\nu_{0,i}) \sigma_{i}(\nu) {\rm d}\nu$ and $\zeta_{i}(t)= \int^{\infty}_{\nu_{0,i}} (4 \pi J_{\nu}(t)/h\nu) \sigma_{i}(\nu) {\rm d}\nu$ are respectively the time-dependent photo-heating and photoionization rates calculated using the photoionization cross-section atomic data  $\sigma_{i}(\nu)$ \citep{Verner1995,Verner1996} and the radiation field $J_{\nu}(t)$ of the time-evolving ionizing stellar population generated by Starburst99, $\nu$ and $\nu_{0,i}$ the frequency and the ionization frequency, respectively, 
$h$ the Planck constant, 
$\alpha_{i}$  the ionic recombination rate including radiative data \citep{Badnell2006} and dielectronic data   \citep[see references in][]{Gray2015}, and $S_{i}$ is the collisional ionization rates from \citet{Voronov1997}.

\subsection{Boundary and Initial Conditions}

To perform hydrodynamic simulations, we assumed the analytic solutions of the fluid model derived by \citet{Chevalier1985} and extended by \citet{Silich2004} for radiative cooling. Based on these solutions,
we set the time-dependent boundary conditions for the density, velocity, and pressure at the cluster radius $r=R_{\rm sc}$ as $\rho_{\rm sc}(t)= \dot{M}(t)/ [2 \pi  R_{\rm sc}^2  V_{\infty}(t)]$, $u_{\rm sc}(t) = \frac{1}{2} V_{\infty}(t)$, and $P_{\rm sc}(t)=\dot{M}(t)V_{\infty}(t) /(\gamma 8 \pi R_{\rm sc}^2)$, respectively, where 
$V_{\infty}(t) = V_{\infty,0} f_{\rm v}(t)$ and $\dot{M}(t) = \dot{M}_{0} g_{\rm \dot{m}}(t)$ are the time-dependent wind velocity and mass-loss rate, respectively, $V_{\infty,0}$ and $\dot{M}_{0}$ the user-defined wind velocity and mass-loss rate at $t=0$, $f_{\rm v}(t)$ is a dimensionless function associated with the time-evolving wind velocity normalized using the initial wind velocity calculated from the mechanical luminosities and mass-loss rates predicted by Starburst99, and $g_{\rm \dot{m}}(t)$ is a dimensionless function made using the time-evolving mass-loss rates produced by Starburst99 normalized with the initial mass-loss rate. The initial conditions of the density, velocity, and pressure outside the cluster radius at $t=0$ are: $\rho_{0}= \mu m_{\rm p} n_{\rm amb}$, $u_{0} = 0$, and $P_{0}= k_{\rm B} n_{\rm amb} T_{\rm amb}$, where $n_{\rm amb}$ and $T_{\rm amb}$ are the number density and temperature of the ambient medium, respectively, $\mu$ is the mean atomic weight ($\mu=0.61$ for a fully ionized gas), $m_{\rm p}$ the proton mass, and $k_{\rm B}$ the Boltzmann constant.
The ambient temperature  $T_{\rm amb}$ is calculated by \cloudy for a stationary medium with $n_{\rm amb}$.

\subsection{Time-evolving Stellar Feedback}

We used the evolutionary synthesis code Starburst99 \citep{Levesque2012,Leitherer2014} to generate the time-evolving radiation field and stellar feedback for stellar population evolution from 1 to 7 Myr. with an initial total stellar mass of $M_{\star}=2\times10^6$\,M$_{\odot}$ and an IMF with the Salpeter $\alpha= 2.35$ for the stellar masses ranging from 0.5 to 150 M$_{\odot}$, using the rotational Geneva population \citep{Ekstroem2012,Georgy2012} and Pauldrach/Hillier atmosphere \citep{Hillier1998,Pauldrach2001}.
The time-dependent ionizing luminosity $L_{\rm ion}(t)$ and spectrum $J_{\nu}(t)$ computed by Starburst99 were employed by the photo-heating efficiencies $\Gamma_i (t)$ in our hydrodynamic simulations and the photoionization rates $\zeta_{i}(t)$ in our photoionization calculations. The time-dependent mass-loss rate $\dot{M}(t)$ at 0.1\,Myr interval calculated by Starburst99 was also used to gradually modify the mass and energy injection rates -- $q_{m}(t)$ and $q_{e}(t)$ --
in our hydrodynamic simulation while it is running.

\section{Numerical Results}

\subsection{Galactic Superwind Modes}

\begin{figure}
\centering
\includegraphics[width=0.48\textwidth, trim = 0 0 0 0, clip, angle=270]{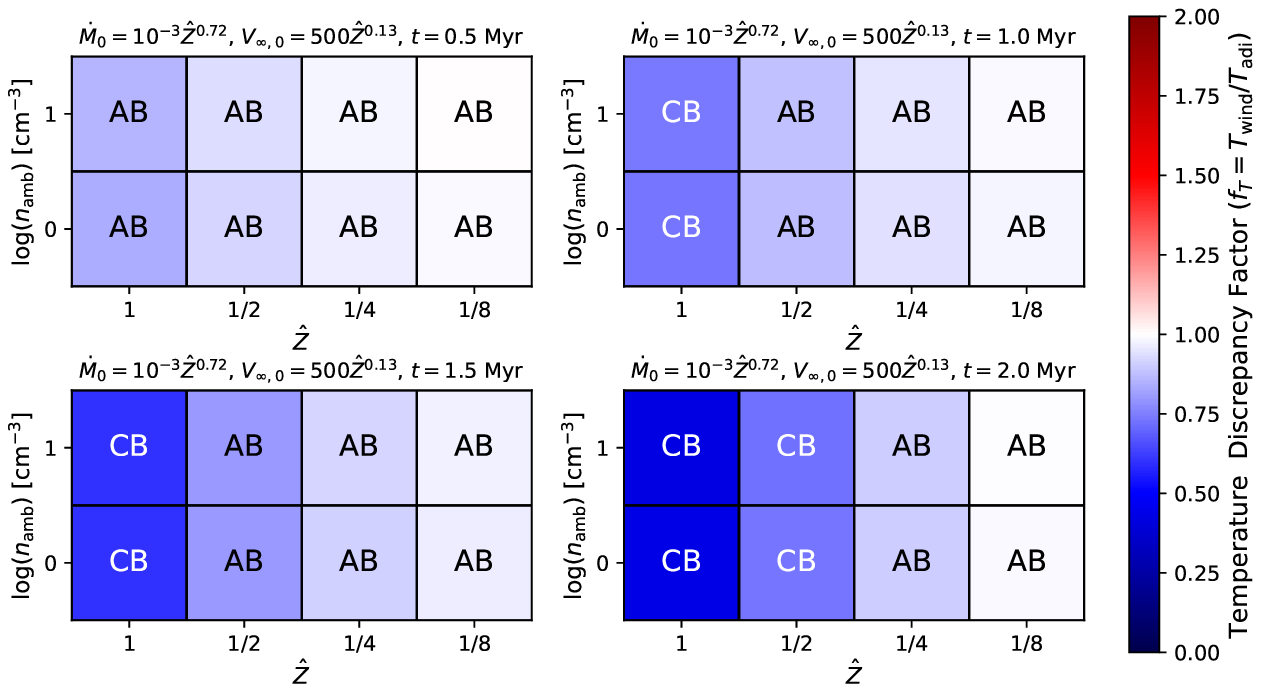}
\caption{The mean wind temperature $T_{\rm wind}$ produced by \maihem with respect to the mean adiabatic solution $T_{\rm adi}$ for the time-dependent wind parameters $V_{\infty}(t)=500 \hat{Z}^{0.13}f_{\rm v}(t)$, km\,s$^{-1}$ and $\dot{M}(t)= 10^{-3} \hat{Z}^{0.72} g_{\rm \dot{m}}(t) $\,M$_{\odot}$\,yr$^{-1}$, metallicities $\hat{Z}\equiv Z$/Z$_{\odot}=[1/8$, 1/4, 1/2, 1], ambient densities $n_{\rm amb}=[1$, 10, 10$^2$, 10$^3]$\,cm$^{-3}$, and a stellar cluster with $R_{\rm sc} =1$\,pc, $M_{\star}=2 \times 10^5$\,M$_{\odot}$, and ages $t=[0.5$, 1, 1.5, 2] Myr.
The wind models are identified as adiabatic bubble (AB) and  catastrophic cooling bubble (CB), based on criteria defined by \citet{Danehkar2021}. 
}
\label{fig2}
\end{figure}

\cite{Danehkar2021} classified galactic superwinds into different wind modes according to the deviation of the wind temperature ($T_{\rm w}$) from the adiabatic solution ($T_{\rm adi}$).
The adiabatic and quasi-adiabatic modes (AW, AB, and AP) are those with mean wind temperatures having $f_T \equiv T_{\rm w} / T_{\rm adi} \geqslant 0.75$. The adiabatic wind (AW) and adiabatic bubble (AB) modes are without and with bubbles, respectively, while the bubble expansion is stalled by the ambient pressure in the adiabatic, pressure-confined (AP) mode. The catastrophic cooling (CC) and catastrophic cooling bubble (CB) modes are those with and without bubbles, respectively, but with $f < 0.75$, while the cooling, pressure-confined (CP) mode describes radiatively cooling with a stalled bubble. Additionally, the no wind (NW) and momentum-conserving (MC) modes describe suppressed superwinds, which were controlled by high ambient pressures and substantial cooling effects, respectively.

Figure~\ref{fig2} presents different wind modes in the space parameters of the ambient density $n_{\rm amb}$, metallicity $Z$/Z$_{\odot}$, and age ($t$) with the time-evolving wind velocity $V_{\infty}(t)=500 \hat{Z}^{0.13}f_{\rm v}(t)$, km\,s$^{-1}$ and mass-loss rate $\dot{M}(t)= 10^{-3} \hat{Z}^{0.72} g_{\rm \dot{m}}(t) $\,M$_{\odot}$\,yr$^{-1}$, 
where the stellar cluster has a radius of $R_{\rm sc} =1$\,pc and a total mass of $M_{\star}=2 \times 10^5$\,M$_{\odot}$. 
We see the formation of radiative cooling in older ages in time-evolving models, 
while higher metallicities and weaker wind velocities contribute to stronger radiative cooling. However, the formation of a bubble cannot always be suppressed by cooling effects, so we have several superwinds in the CB mode \citep[see also Fig. 4 in][]{Danehkar2021}.

In Figure~\ref{fig1}, the temperature and density profiles (solid red lines) of a superwind predicted by our hydrodynamic simulation are plotted in the left panels against the adiabatic solutions (red dashed lines). 
The profiles are divided into four different regions according to \citet{Weaver1977}, namely (1) wind, (2) bubble, (3) shell, and (4) ambient medium  (see dotted, dashed, and dash-dotted gray lines). 
The Str\"{o}mgren sphere (solid gray line) was also shown, which is predicted by a pure photoionization model.

\subsection{Collisional Ionization versus Non-equilibrium Ionization}

\begin{figure}
\centering
\includegraphics[width=0.48\textwidth, trim = 0 0 0 0, clip, angle=270]{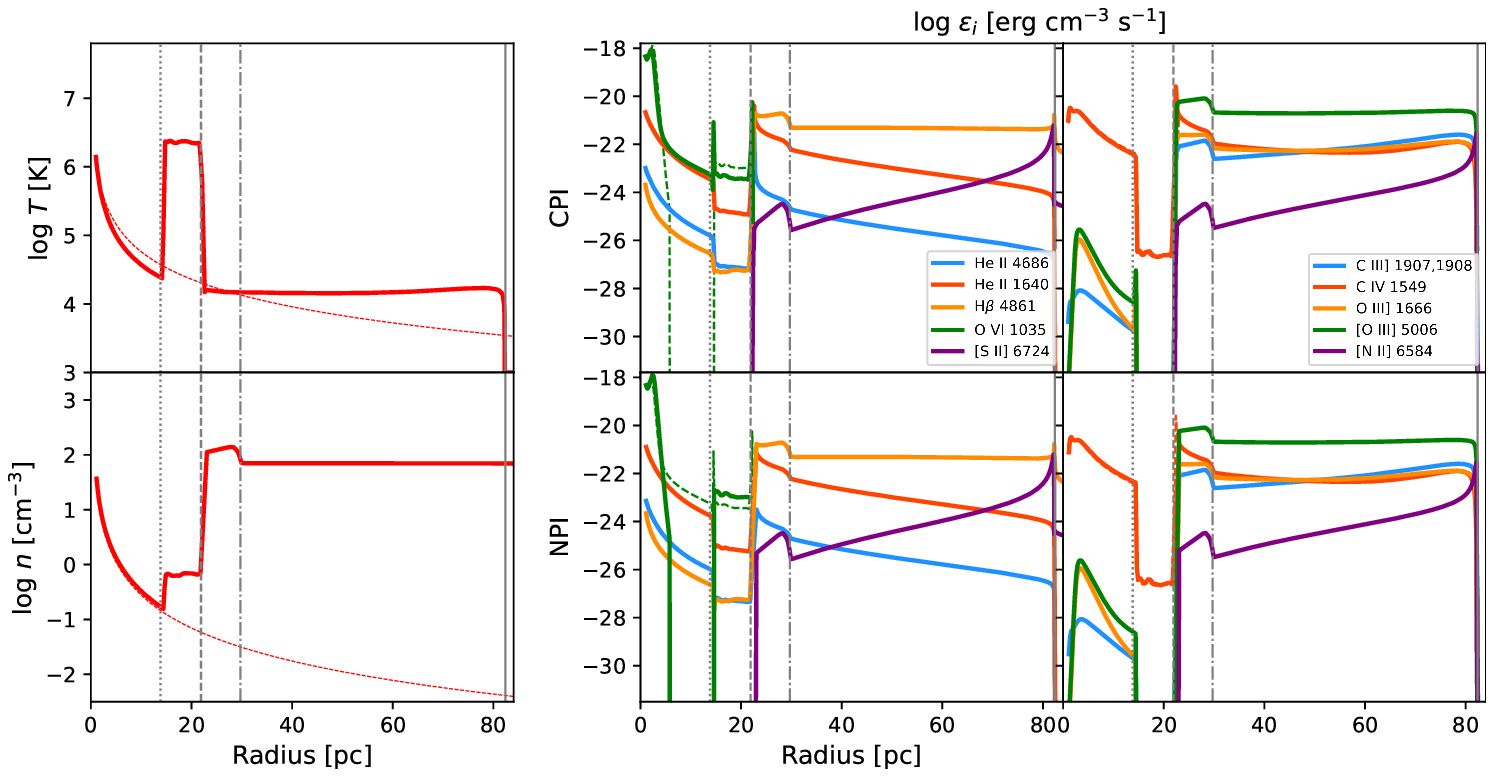}
\caption{The temperature ($T$) and density ($n$) profiles (solid red lines) plotted against the adiabatic solutions (red dashed lines). The logarithmic emissivities $\log \varepsilon_{i}$ of different emission lines predicted by hybrid collisional ionization and photoionization (CPI) and non-equilibrium photoionization (NPI) calculations.
The boundaries of the bubble, shell end, and Str\"{o}mgren sphere are split using dotted, dashed, dash-dotted, and solid gray lines. The model parameters are: $V_{\infty}=418$\,km\,s$^{-1}$, $\dot{M}= 0.369 \times 10^{-2}$\,M$_{\odot}$\,yr$^{-1}$, $t=1$\,Myr, $R_{\rm sc} =1$\,pc, $M_{\star}=2 \times 10^6$\,M$_{\odot}$, $n_{\rm amb}=100$\,cm$^{-3}$, $Z/Z_{\odot}=0.25$.
The \ionic{O}{vi} lines predicted by NPI are overplotted by dashed green lines in the CPI panel, and vice versa.
}
\label{fig1}
\end{figure}

To create collisional ionization, \citet[][]{Danehkar2021} employed the density and temperature profiles produced by our hydrodynamic simulations, along with the radiation field, to calculate the emission-line profiles with the photoionization code \cloudy \citep{Ferland2017}. Our hydrodynamic simulations also generate time-dependent ionization states using Eq.\,(\ref{eq_4}), which can make non-equilibrium ionization (NEI) in the cooling ($<10^{6}$\,K) regions where collisional ionization takes longer than radiative cooling. To produce non-equilibrium photoionization,  \citet[][]{Danehkar2022} performed zone-by-zone \cloudy computations by running one individual photoionization model for the given temperature, density, and NEI states of each zone, while each hydrodynamic simulation typically contains 1024 zones. Pure photoionization is applied to the ambient medium.

Figure~\ref{fig1} shows the emissivity profiles of different emission lines predicted by hybrid collisional ionization and photoionization (CPI; top-right panels) and non-equilibrium photoionization calculations (NPI; bottom-right panels) made with the physical properties and time-dependent ionization states produced by our hydrodynamic simulations \citep[for a wider parameter range, see Fig.\,3 in][]{Danehkar2022}. It can be seen that the \ionic{O}{vi} 
emission line does not have the same emissivity profile in the NPI status as  it does in the CPI status, especially at temperatures below $10^{6}$\,K in the radiative cooling region. This effect is a consequence of time-dependent photoionization, which occurs when the radiative cooling timescale is much faster than the collisional ionization timescale.

\section{Applications in Starburst Regions}

\citet{Danehkar2022} found that the 
\ionic{O}{vi} $\lambda\lambda$1032,1038 emission-line fluxes predicted by non-equilibrium photoionization demonstrate noticeable enhancements in metal-poor models. 
As proposed by \citet{Gray2019a,Gray2019}, the enhanced \ionic{O}{vi} lines could be linked to strong radiative cooling in starburst regions. 
\ionic{O}{vi} emission with a velocity offset of about $50$--100\,km\,s$^{-1}$ in the rest frame was detected toward a hot bubble in the nearby spiral galaxies NGC\,4631, providing evidence for gas cooling \citep{Otte2003}.
For the intense starburst J1156+5008 at $z=0.236$ with an \ionic{O}{vi} absorption blueshifted by 380 km\,s$^{-1}$, \citet{Hayes2016} concluded that the \ionic{O}{vi}-carrying gas must be cooling in situ via the coronal phase. Moreover, \citet{Chisholm2018} proposed the creation of \ionic{O}{vi} absorption in the high-redshift ($z=2.92$) galaxy J1226+2152 by either the conductive evaporation of cool gas or a cooling flow between a hot outflow and a cooler photoionized gas.

The nearest ($d=82$\,Mpc), Lyman-break analog Haro\,11, also depicts \ionic{O}{vi} $\lambda\lambda$1032,1038 absorption features with a wind velocity of 200--280 km\,s$^{-1}$, which could be attributed to radiative cooling superwinds 
\citep{Grimes2007}. \ionic{O}{vi} emission was also identified in Haro\,11, for which \citet{Grimes2007} estimated that up to 20\% of the supernova feedback was lost to possible radiative cooling. 
It is one of the nearest analogues to high-redshift galaxies because of its high star formation and relatively low metallicities. Haro\,11 includes three main knots (A, B, and C) 
consisting of several super star clusters with a cluster age distribution up to 40 Myr and a peak age of 3.5 Myr \citep{Adamo2010}.
While Knot B could produce energy-driven superwinds due to the presence of visible bubbles, the lack of any bubbles around Knot C may be an indication of momentum conservation or substantial radiative cooling.
The ionized gas around Knot C was found to have a metallicity of 0.12\,Z$_{\odot}$, a factor of 3 lower than 
the ISM in Knot B \citep{James2013}, so the metallicity does not seem to have a major role in the formation of momentum-driven or radiatively cooled superwinds in Knot C. 
Future high-resolution UV spectroscopic measurements of different regions of the knots in Haro\,11 will allow us to determine which of them primarily bears \ionic{O}{vi} absorbing winds.
Our hydrodynamic simulations and non-equilibrium photoionization calculations with time-evolving stellar feedback at ages beyond 1\,Myr and wider parameter ranges of superwinds and star clusters will certainly improve our understanding of observed \ionic{O}{vi} lines and their possible implications for radiative cooling in starburst regions.






\end{document}